# Surface Plasmon-Coupled Enhanced Transmission

**Amir Djalalian-Assl** [1]

[1]51 Golf View Drive, Craigieburn, VIC 3064, Australia

*Correspondence: amir.djalalian@gmail.com

**Abstract:** For distances less 10 nm, a total energy transfer occurs from a quantum emitter to a nearby metallic surface, producing evanescent surface waves that are plasmonic in nature. When investigating a metallic nanohole supported on an optically dense substrate (such as diamond with nitrogen vacancy center), the scattering occurred preferentially from the diamond substrate towards the air for dipole distances less 10 nm from the aperture. In addition, an enhancement to the dipole's radiative decay rate was observed when resonance of the aperture matched the emitters wavelength. The relationship between an emitter and a nearby resonant aperture is shown to be that of the resonance energy transfer where the emitter acts as a donor and the hole as an acceptor. In conjunction with the preferential scattering behavior, this has led to the proposed device that operates in transmission mode, eliminating the need for epi-illumination techniques and optically denser than air superstrates in the collection cycle, hence making the design simpler and more suitable for miniaturization. A design criterion for the surface grating is also proposed to improve the performance, where the period of the grating differs significantly from the wavelength of the surface plasmon polaritons. Response of the proposed device is further studied with respect to changes in nitrogen vacancy's position and its dipolar orientation to identify the crystallographic planes of diamond over which the performance of the device is maximized.

---

## 1. Introduction

The use of optically dense materials to increase the collection efficiency associated with a quantum emitter (QE) is popular[1-8]. The inclusion of diamond superstrates[4] with its low absorption[9], for example, allows the emission from a Nitrogen Vacancy center(NV-) in a nano-diamond to be readily collectable from the superstrate. Most designs, however, are based on the epi-illumination techniques where the incident pump power and the luminescent radiation share the same half-space. See for example the study undertaken by Aouani *et. al.* [10,11] on the fluorescence of an ensemble of molecules with randomly oriented dipole moments, trapped inside a plasmonic aperture, when illuminated and collected through the glass substrate. A more deterministic study showed the dependence of the radiation pattern of plasmonic antennas with respect to the distance and the dipolar orientation of a nearby nano-diamond with NV- [7,12].

Devices based on reflection geometry rely on point-illumination and point-collection using the same objective lens which is an intrinsic feature of the confocal microscopy. Maintaining an optimal distance between the objective lens and the emitter, therefore, is a crucial factor in maximizing the incident power over a single photon emitter. Moving the objective lens away from the emitter into the far-field zone for collection purposes, (hence moving the emitter away from the focal point), reduces the incident pump power over the dipole drastically. Consequently, the excitation of a dipole in reflection mode must either be carried out by side-illumination techniques via a separate objective lens, or by a confocal lens in conjunction with a Bertrand lens allowing for collection from the back focal plane. Such configurations besides being more complex require extra elements on the optical path. For all miniaturization purposes, therefore, it is beneficial for a device to operate in transmission mode where the incident pump utilized the substrate and luminescence is collected from the superstrate. In the context of detecting a quantum emitter, however, this implies that a device must possess multiple qualities, such as enhancement and collimation of the emission.

Recent developments in nano-diamond fabrications have led to designs such as a diamond film on a glass substrate with bullseye (BE) grating etched on the diamond surface surrounding a nitrogen vacancy to collimate its emission[13]. The use of a homogenous diamond film with NV- near the surface aimed to minimize complications (such as total internal reflection) associated with alternative designs where nano-diamonds are integrated with external structures. Although such a multilayer structure may operate in either the reflection or the transmission mode, numerical analysis showed a higher field intensity being trapped inside the diamond film in comparison to those scattered into the air or into the glass substrate[13]. This may be attributed to the complex interaction of the NV- emission with the air/diamond and/or diamond/glass boundaries, which includes total internal reflections and the quenching of the dipole's emission by nearby *dielectric* interfaces. In this report, it will be shown that by introducing a nearby *metallic* surface, it is possible to prevent the NV-'s emission from being trapped or scattered inside the diamond membrane.

Understanding the interaction between a dipole and nearby metallic surfaces has led to the Surface Plasmon-Coupled Emission (SPCE)[14]. Theoretical and experimental studies have shown that fluorophores within ~10 nm of metallic surfaces are quenched[1]. The quenching effect, however, is understood based on analytical solutions and experimentally reported data being focused on the reflected power or the decayed power through the flat metallic thin films [14-19]. The quenching region of less than ~10 nm away from metallic surfaces is in fact related to the total energy transfer between the dipole that acts as a donor and the metallic surface acting as an acceptor [20]. The quenching of quantum emitters by a nearby *metallic* surface results in surface waves that propagate at the supercritical angle along the metal/dielectric interface[21], not being detectible in the far-field. This mechanism is advantageous if utilized properly. It can be shown that such evanescent surface waves are plasmonic in nature and may be utilized favorably. In other words, by positioning a dipole at distances of ~10 nm (or less) from metals, it is possible to consume most of the power emanating from the dipole to induce Surface Plasmon Polaritons (SPP) that propagate over the bulk metallic surfaces and/or excite Localized Surface Plasmons (LSP) in subwavelength metallic features such as a groove or an aperture. A preliminary study of a resonant aperture in a silver film on a diamond substrate with NV- 10 nm away from the aperture, [22], showed the emission from the aperture occurred preferentially towards the air, meaning that evanescent surface waves and freely propagating EM waves were much stronger in the air half-space in comparison to those inside the diamond. This is partially due to the SPPs having longer decay length of ~67 μm at the silver/air interface in comparison to ~3.2 μm at the silver/diamond interface. Such a behavior provided an opportunity for a new design.

The platform propose here consists of a single crystal diamond membrane with a NV- near its surface[9,23-25] that acts as a donor, integrated with a planar plasmonic antenna acting as an acceptor, to achieve a highly unidirectional transmitted luminescence with enhanced Radiative Decay Rate (RDR) and high collection efficiency in the air, hence, eliminating the need for optically dense superstrates and epi-illumination. In the case of a diamond substrate, being a material of choice in this report, it is equally important to identify optimum crystallographic planes of diamond over which the proposed planar structure to be fabricated. With NV-'s symmetry axis being one of the <111> crystallography directions[26,27], it is possible to identify such planes in which the energy transfer efficiency from NV- to the antenna is maximized. Such a platform may also be realized by other substrates with near surface emitters such as a SiC with defects [28] or a silicon with embedded quantum dots [29]. A review on quantum dots and ways to enhance its emission is given by Xu *et al.*[30]

## 2. Resonance Energy Transfer

A rigorous theoretical study on dipolar emission near a flat metallic surface was previously carried out by Novotny [21,31]. Here a numerical study of a diamond membrane with a NV- positioned at $z_d$ is undertaken, where $z = 0$ plane marks the diamond/air interface. The model is developed in 2D due to the limited computational resource available to the author. The analysis is then extended to a multilayer diamond/silver/air structure. Finite Element Method (FEM) simulations are carried out at the target wavelength of $\lambda_0 = 700$ nm that coincides with the spectral peak of the phonon sideband associated with the emission from nitrogen vacancies in nano-diamonds at room temperature [32,33]. The refractive index of the diamond substrate is set to $n_{sub} = 2.4$ and the refractive index data for silver was taken from Palik [34]. Figure 1(a) depicts the scattered electric field intensity, $|E|^2$, calculated for a diamond membrane with a NV- positioned at $(x_d, z_d) = (0, -10)$ nm with its dipole moment aligned with the $x$-axis (i.e. horizontally oriented). The power ratio of $P_{air}/P_{sub} \approx 0.5$ is a clear indication that in the absence of any plasmonic structure most of the emitted power is scattered into the diamond. The power ratio $P_{air}/P_{sub} = \int \langle \mathbf{S} \rangle_{air} \cdot d\mathbf{s} \Big/ \int \frac{\langle \mathbf{S} \rangle_{sub}}{n_{sub}} \cdot d\mathbf{s}$ is calculated over a closed arc **s** where $\langle \mathbf{S} \rangle_{air,sub}$ are the time averaged Poynting vectors calculated in the air and inside the diamond substrate respectively. Introducing an optically thick silver film on top of the diamond substrate reduced the scattered power into the air to ~0, see Figure 1(b). Electric field components, $(E_x, E_z)$, calculated over the diamond/silver interface, Figure 1(e), with their corresponding Fast Fourier Transforms (FFT), Figure 1(g), revealed the presence of surface waves, having a wavelength $\lambda_{SPPd} = 230$ nm. This agrees with the Surface Plasmon Polaritons (SPP) wavelength $\lambda_{SPPd} = 246$ nm obtained analytically using $k_{SPP} = \mathrm{Re}\left[\sqrt{\frac{\varepsilon_m \varepsilon_d}{\varepsilon_m + \varepsilon_d}} k_0\right]$ at $\lambda_0 = 700$ nm, where $k_0$ is the free space wavenumber, $\varepsilon_m$ and $\varepsilon_d$ are relative permittivities of the metal and dielectric respectively. It is apparent that some of the emanating power from the dipole is consumed in launching SPPs. Presence of a dipole-like activity associated with a metallic slit has been reported previously [35]. Therefore, by introducing a resonant aperture perforating the film just above the NV-, it is possible establish the dipole-dipole *resonance energy transfer* between the NV- and the aperture.

Figure 1(c) shows the enhancement to the Radiative Decay Rate ($\mathrm{RDRE}_{air}$) in the air half-space vs. the film thickness, $t$, when an aperture having a width $w_a = 30$ nm, perforates the silver film just above the emitter. The $\mathrm{RDRE}_{air}$ was calculated for the air half-space using $\mathrm{RDRE}_{air} = P_{air}/(0.5 \times P_0) = \int \langle \mathbf{S} \rangle_{air} \cdot d\mathbf{s} \Big/ 0.5 \times \int \langle \mathbf{S} \rangle_{vac} \cdot d\mathbf{s}$, where $P_0$ is the total power emitted by a nano-diamond in vacuum. The diameter of the isolated nano-diamond was also set to 30 nm in all calculations. This is a typical size for a nano-diamond with luminescence properties [36-38]. The factor of 0.5 in the denominator is due to $P_0$ being calculated along the arc length of a full circle, whereas $P_{air}$ was calculated over a semicircle in the air half-space. With $z_d = -10$ nm, a maximum $\mathrm{RDRE}_{air} \approx 140$ was achieved for $t = 110$ nm that corresponds to the first Fabry-Pérot resonance of the aperture [39]. Figure 1(d) depicts the electric field intensity, $|E|^2$, calculated for the diamond/silver/air multilayer with a 30 nm wide aperture perforating the 110 nm silver film just above the NV-. The power ratio $P_{air\text{-}Ag110nm+hole}/P_{sub\text{-}Ag110nm+hole} = 10$ is a clear indication that scattered power is preferential towards the air. Calculated energy transfer efficiency between the donor and the acceptor [21], $E_T = P_{D\rightarrow A}/(P_D + P_{D\rightarrow A}) = P_{air\text{-}Ag110nm+hole}/(P_{air\text{-}noFilm} + P_{air\text{-}Ag110nm+hole}) \approx 1$ was obtained for $z_d \leq -10$ nm. Power ratios $P_{sub\text{-}Ag110nm+hole}/P_{sub\text{-}noFilm} = 1.5$ and $P_{air\text{-}Ag110nm+hole}/P_{air\text{-}noFilm} = 33$, however, suggest that the presence of the silver film with a resonant aperture enhances the reflected power inside the diamond as well as the transmitted power into the air, hinting at a change in the density of states. It is therefore intuitive to infer that $\mathrm{RDRE}_{air} \propto P_{D\rightarrow A}$. Radiation pattern of transmitted field, however, is dispersive with $1/e$ of its maximum intensity not reaching beyond $z \approx 1.7$ μm along the optical axis, i.e. $z$-axis. Figure 1(d) also shows the presence of high intensity evanescent field along the silver/air interface. Electric field components, Figure 1(f), and their corresponding FFTs, Figure 1(h), calculated over the silver/air interface, confirmed the presence of surface waves with $\lambda_{SPPa} = 667$ nm which is in agreement with $\lambda_{SPPa} = 682$ nm obtained analytically. Although evanescent waves do not contribute to the far-field [21], it is possible to intercept and scatter them into freely propagating EM waves by introducing periodic surface gratings surrounding the aperture [40]. Note however, that for a horizontally oriented dipole moment, induced $E_x$ is an even function of $x$ whereas $E_z$ is an odd function. Consequently, any contribution made by $E_z$ to the scattered fields leads to a destructive interference along the optical axis. Therefore, considering a 2D model, $E_x$ (with even parity) is the only contributing factor to the far-field intensity along the optical axis.

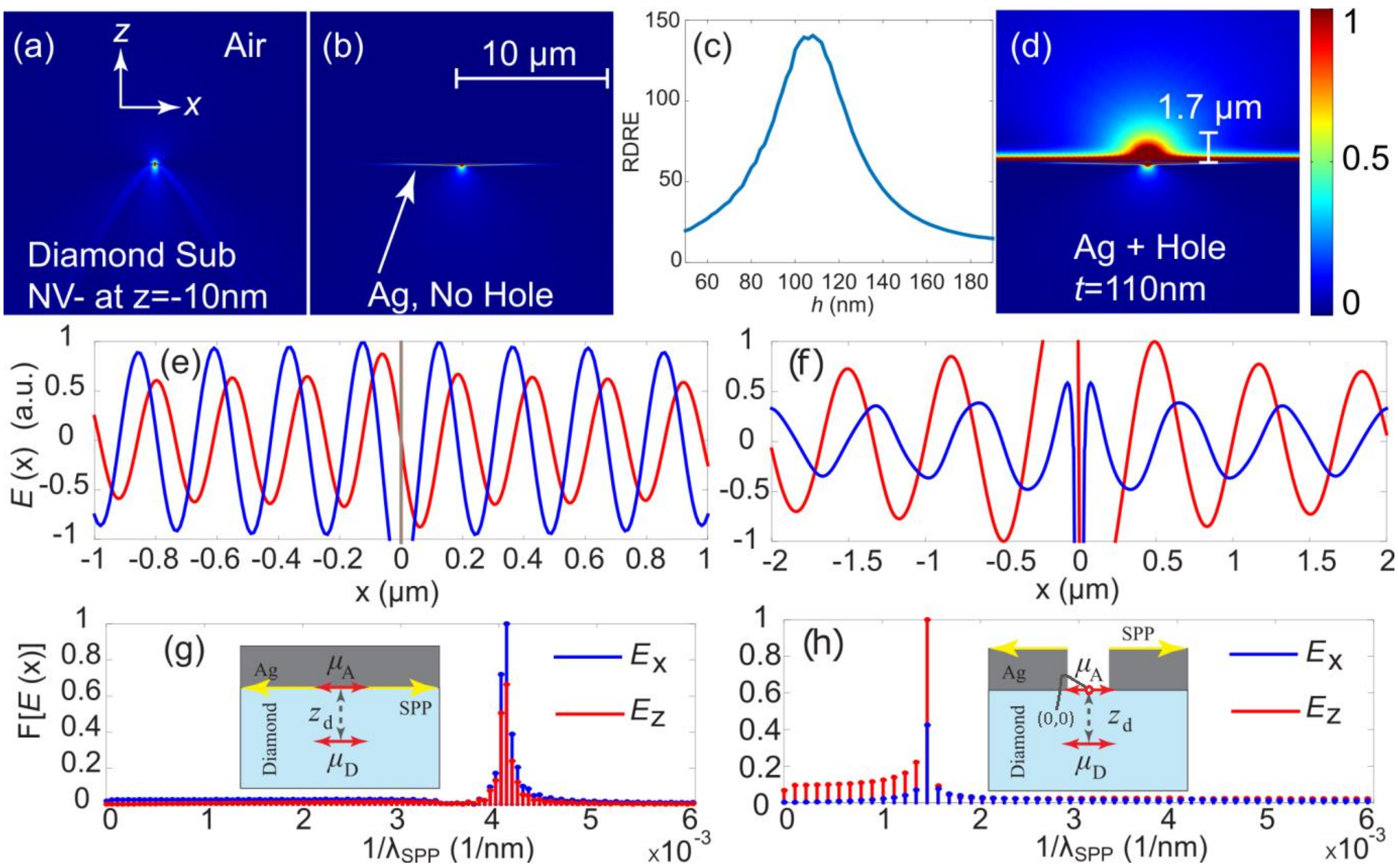

Figure 1. $|E|^2$ calculated for a diamond substrate with NV- with horizontally oriented dipole moment, positioned at ($x_d$, $z_d$) = (0, -10) nm (a) with no metallic film and (b) with an optically thick silver film over the diamond substrate. (c) $RDRE_{air}$ vs. the film thickness with a 30 nm wide aperture perforating the film above the NV-. (d) $|E|^2$ calculated for diamond/silver/air with a 30 nm wide aperture perforating the 110 nm silver film above the NV-. $E_x$ and $E_y$ calculated at (e) diamond/silver with no aperture and (f) silver/air interfaces with the aperture. Fast Fourier transforms of $E_x$ and $E_y$ calculated at (g) diamond/silver and (h) silver/air interfaces.

## 2. Design of Surface Gratings

Detecting the transmitted luminescence of a QE through the aperture, requires reshaping its dispersed radiation pattern into a highly directional light with a maximum possible intensity along the optical axis of a BE structure. Tailoring the exit surface surrounding the aperture with properly designed corrugations is crucial since poor choices may lead not only to the suppression of an already enhanced RDR, but also the formation of side-lobes in the radiation pattern, hence a reduction of transmitted power along the optical axis. Analytical models based on forward propagating SPPs has led to discrepancies between the model and the experimentally observed shortened propagation length attributed to excess losses that required experimentally obtained fitting parameter to correct the theory[41]. It can be shown, however, the shortening of the propagation length is due to the superposition of surface waves. Relation governing the EM waves along the optical axis of a BE structure consisting of a hole surrounded by $N$ periodic corrugations with periodicity $P$, may be described by:

$$\Psi\big|_{x=0} = \psi_{\mathrm{h}} + \sum_{n=1}^{N} \psi_n \quad (1)$$

where $\psi_{\mathrm{h}} = A_{\mathrm{h}} \dfrac{e^{ik_0 z}}{z}$ is the wave with an amplitude $A_h$ emanating from the hole in the $z$-direction, $\psi_n = A_n \dfrac{e^{i(k_0 r_n + \phi_n)}}{r_n} \sin\theta_n$ is the scattered wave by the $n^{th}$ groove having an amplitude $A_n$, $r_n = \sqrt{(nP)^2 + z^2}$ is the distance between $n^{th}$ groove and an arbitrary point along the $z$-axis, $\phi_n = \angle \Psi_{SPP}\big|_{x=x_n}$ is the phase of the superposed (not just forward propagating) SPP waves at the $n^{th}$ groove, see Figure 2(a). The role of a corrugation (or a groove) positioned at $x_n = nP$, is to utilize some of the power carried by the SPPs to excite LSPs inside the groove, Figure 2(b), whose far-field radiation pattern is similar to that of a dipole antenna. The strength of the LSP inside the $n^{th}$ groove is denoted by $A_n$. To simplify the argument consider equation (1) with $N = 1$ in a symmetric configuration consisting of an aperture at $x = 0$ and a pair of grooves at $x_1$ and $x_{-1}$ with $x_{-1} = - x_1$, where the superposition of SPPs at $x = x_1$ may be defined as:

$$\Psi_{\mathrm{SPPa}}\big|_{x=x_1} = e^{ik_{\mathrm{SPP}} x_1} + e^{i[k_{\mathrm{SPP}}(3x_1)+\pi]} + e^{i[k_{\mathrm{SPP}}(2x_1)+2\pi]} \qquad (2)$$

Here, the first term on the right-hand side is the forward wave launched by the aperture arriving at $x_1$, the second term is the wave launched by the aperture towards $x_{-1}$ being reflected back towards $x_1$, and the third term is the reflected wave from $x_1$ towards the aperture, being reflected back towards itself. The phase difference between the wave emanating from the aperture and the wave scattered by the groove calculated along the $z$-axis, $\Phi\big|_{x=0} = \angle\psi_{\mathrm{h}} - \angle\psi_1$ vs. $x_1$, showed that for $x_1 = m\lambda_{SPP}/2$, $\Phi \rightarrow 0$ asymptotically only for $z \rightarrow \infty$, Figure 2(c), but $z = \infty$ is not a quantitative measure for the far-field. Although matching the period of the corrugations to the SPP wavelength leads to a collimated beam at $z = \infty$, simulations confirmed that it did not produce the highest possible intensity along the $z$-axis (results not shown here). Therefore, matching the period of the corrugations to $\lambda_{SPPa}$ is an oversight.

An alternative criterion for maximizing the field intensity along the optical axis is proposed here. Figure 2(d) depicts the square of the amplitude of the superposed SPP wave, $\left|\Psi_{\mathrm{SPPa}}\right|^2_{x=x_1}$, vs. $x_1$ obtained at the silver/air interface using equation (2). Note that the square of the amplitude is proportional to the total power available to excite LSPs inside the groove, hence $A_1 \propto \left|\Psi_{\mathrm{SPPa}}\right|_{x=x_1}$. This simple configuration was also modelled in 2D using FEM. The numerically obtained magnetic field intensity, $\iint \left|H_y\right|^2_{x=x_1} dxdz$, calculated inside the groove vs. $x_1$ agrees with the analytical solution. Total power available to the groove follows a Fabry-Pérot like resonance that depends on the groove's distances from the aperture. The strongest LSP excitation was observed when the groove was positioned at $x_1 = 505$ nm. The situation becomes further complicated when considering the impact of Fabry-Pérot oscillations on the total power inside aperture, hence the enhancement/suppression of the $RDR_{air}$. Figure 2(d)-(inset) shows that $\iint \left|H_y\right|^2_{x=0} dxdz$ calculated inside the aperture has its maximum when $x_1 = 475$ nm and well away from $x_1 = \lambda_{SPPa}$, marked by a star.

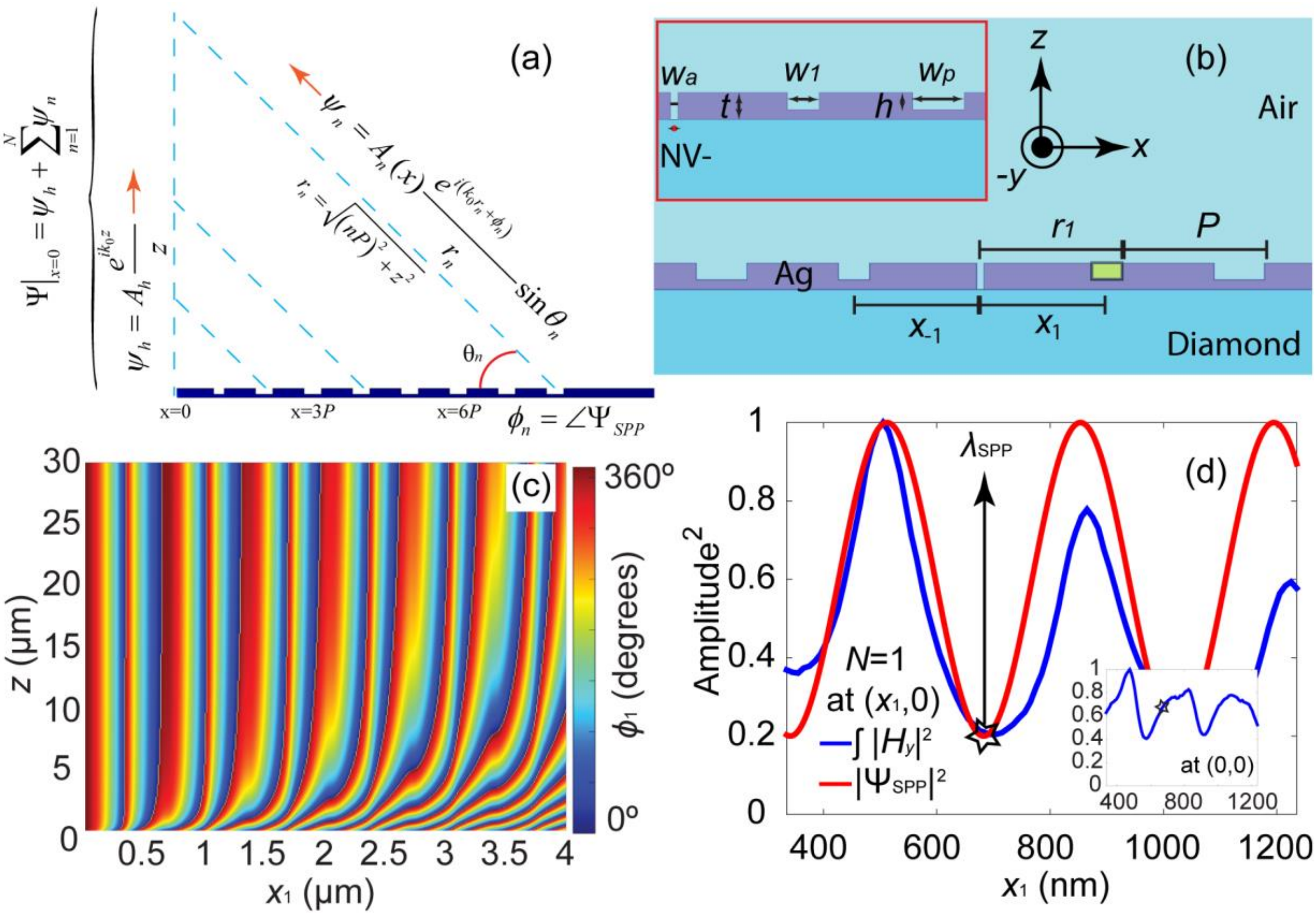

Figure 2. (a) Wave components governing the interference mechanism along the optical axis of a plasmonic bullseye device. (b) Schematic of a bullseye device and its components. (c) The phase difference between the wave emanating from an aperture and the wave scattered by a groove, $\Phi|_{x=0} = \angle\psi_{\mathrm{h}} - \angle\psi_1$, vs. $x_1$ calculated along the $z$-axis bases on superposed SPPs. (d)- Square of the amplitude of superposed surface waves, $|\Psi_{\mathrm{SPPa}}|^2_{x=x1}$, obtained analytically and $\iint |H_y|^2_{x=x_1} dxdz$, obtained numerically. (inset) $\iint |H_y|^2_{x=0} dxdz$ calculated numerically inside the aperture.

Given that the aperture and the grooves form an antenna array, as long as their equidistance is less than $\lambda_0$, high intensity radiation along the optical axis is guaranteed [42] (p. 399). But one must also maximize the radiating power available to both the aperture and the grooves. It is intuitive, therefore, to set $x_1$ to the average of the two peaks mentioned above, i.e. $x_1 = 0.5 \times (475 + 505)$. It is obvious that equation (2) becomes convoluted for $N > 1$. Considering such complex interactions, the only viable approach in designing a BE structure in the presence of a quantum emitter, is to solve the Maxwell equations numerically. With $t = 110$ nm and $w_a = 30$ nm being the optimum values for the film thickness and the aperture width, $r_1$ was set to $x_1 + w_1/2 = 555$ nm and a series of parametric sweeps were carried out over each of the remaining parameters $w_1$, $w_p$, $h$ and $P$ for $z_d = -10$ nm. In each case, the spectrum for the scattered power into the air vs. the parameter was calculated to identify the peak position that corresponds to the optimum dimension. Note that this is the same approach used in determining the optimum film thickness, $t$, see Figure 1(c). One possible configuration, C1, capable of collimating the emission as well as improving the $\mathrm{RDR}_{\mathrm{air}}$ was found to be $w_a = 30$ nm, $w_1 = 130$ nm, $w_p = 210$ nm, $h = 70$ nm, $r_1 = x_1 + w_1/2 = 555$ nm, $P = 590$ nm and $t = 110$ nm. Enhancement to the transmitted $\mathrm{RDR}_{\mathrm{air}}$ peaks for $N = 5$ with no significant changes to the radiation pattern for $N > 5$, see Figure 3. This implies that most (if not all) of the power carried by the SPPs are intercepted and scattered by the first five grooves. Therefore, for the rest of the calculations only 5 grooves are considered. Note that in this report, all 2D field plots are normalized to the same scale and comparable to one another.

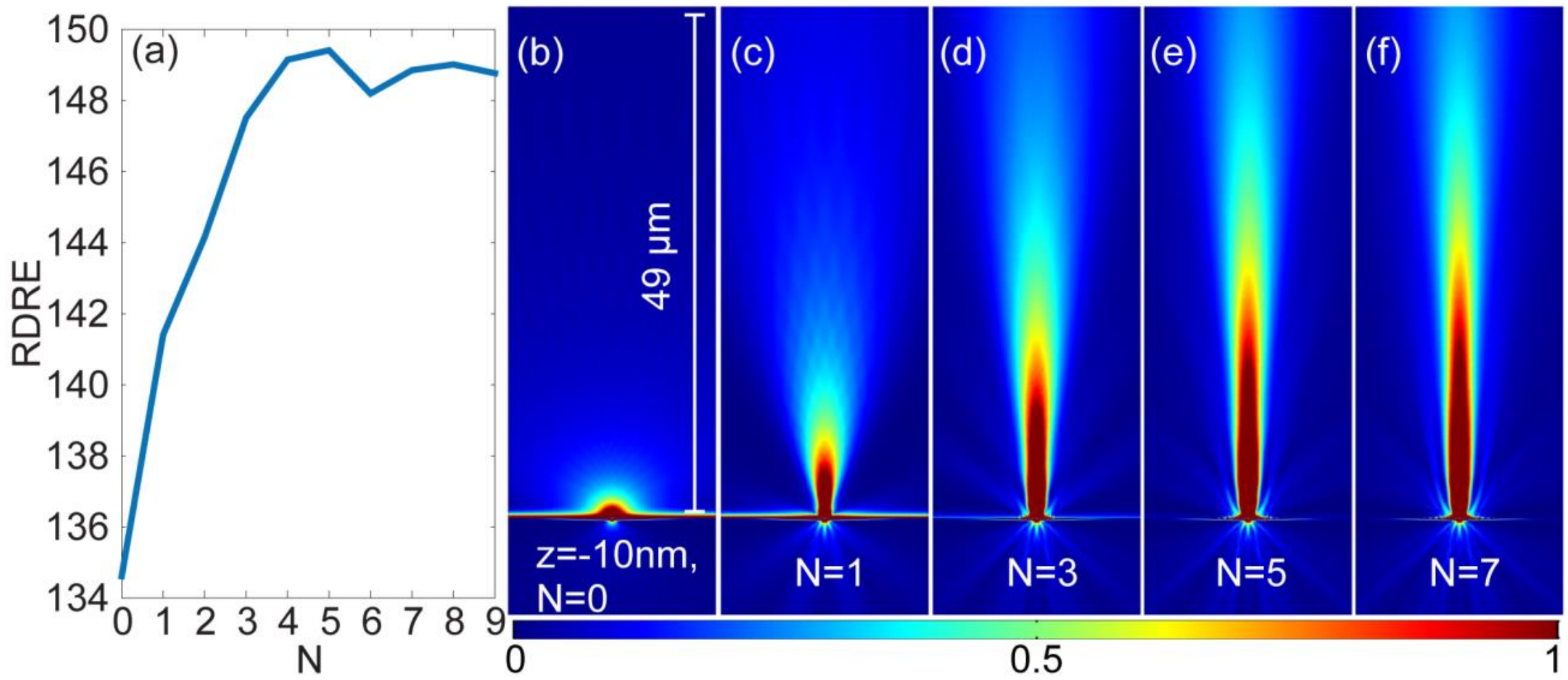

Figure 3. (a) $RDRE_{air}$ and (b)-(f) $|E|^2$ vs. the number of grooves $N$ for a BE with $r_1$ = 555 nm, $P$ = 590 nm, $w_1$ = 130 nm and $w_p$ = 210 nm, $w_a$ = 30 nm and $t$ = 110 nm, on a diamond substrate with NV- at ($x_d$, $z_d$) = (0,-10) nm, emitting at $\lambda$ = 700 nm.

With the smallest spot size achievable by an Electron Beam Lithography (EBL) or a Focused Ion Beam (FIB) being as low as ~5 nm[43,44], fabrication of the aperture and the surrounding surface features poses no challenge. The resonance energy transfer between the donor and the acceptor, on the other hand, is sensitive to the changes in dipole's position and orientation. The rest of the report, therefore, is dedicated to study such variations.

## 3. Dipole's Position and Orientation

The transmitted collection efficiency CE = $P_{air\text{-}C1}/(P_{air\text{-}C1} + P_{sub\text{-}C1})$ and $RDRE_{air}$ vs. $z_d$ is depicted in Figure 4(a). Note that preliminary investigations with 1 nm steps, revealed the presence of a single peak in $RDRE_{air}$ at $z_d$ = -3 nm. Therefor the resolution for the final parametric sweep over $z_d$ was set to 1 nm in the range of -5 nm ≤ $z_d$ ≤ 0 and 5nm for $z_d$ < -5 nm. An enhancement of $RDRE_{air}$ ≥ 216 was achieved for $z_d$ ≤ -5 nm with a maximum of 219 at $z_d$ = -3 nm and a drop to $1/e$ at $z_d$ = 17 nm. Rate of change in CE vs. $z_d$ was found to be much slower in comparison to that of $RDRE_{air}$. A minimum of 80% efficiency in the range of $z_d$ ≤ 25 nm with a maximum of 86% at $z_d$ = -5 nm was obtained. For of a horizontally oriented dipole, transmitted $RDRE_{air}$ showed a reciprocal relation to those corresponding to dipole emissions near flat metallic surface, compare Figure 4(a) in this report to Figure 10.5(b) in "Principles of nano-optics"[21], for example. The slow drop of ~$1/z_d^{0.8}$ in $RDRE_{air}$ is attributed to the energy transfer between the NV- and the resonant aperture that involves a mutli-channel mechanism. While the direct excitation of LSPs inside the resonant aperture, with its finite cross section, is dominated by short range interactions, the surface surrounding the aperture is able to capture NV-'s emission for longer distances, producing SPPs which consequently couple to the aperture's LSPs. This slow rate of change with respect to the dipole's distance from the surface is extremely beneficial. Beaming profiles for few arbitrary $z_d$ values are depicted in Figure 4(b)-(e).

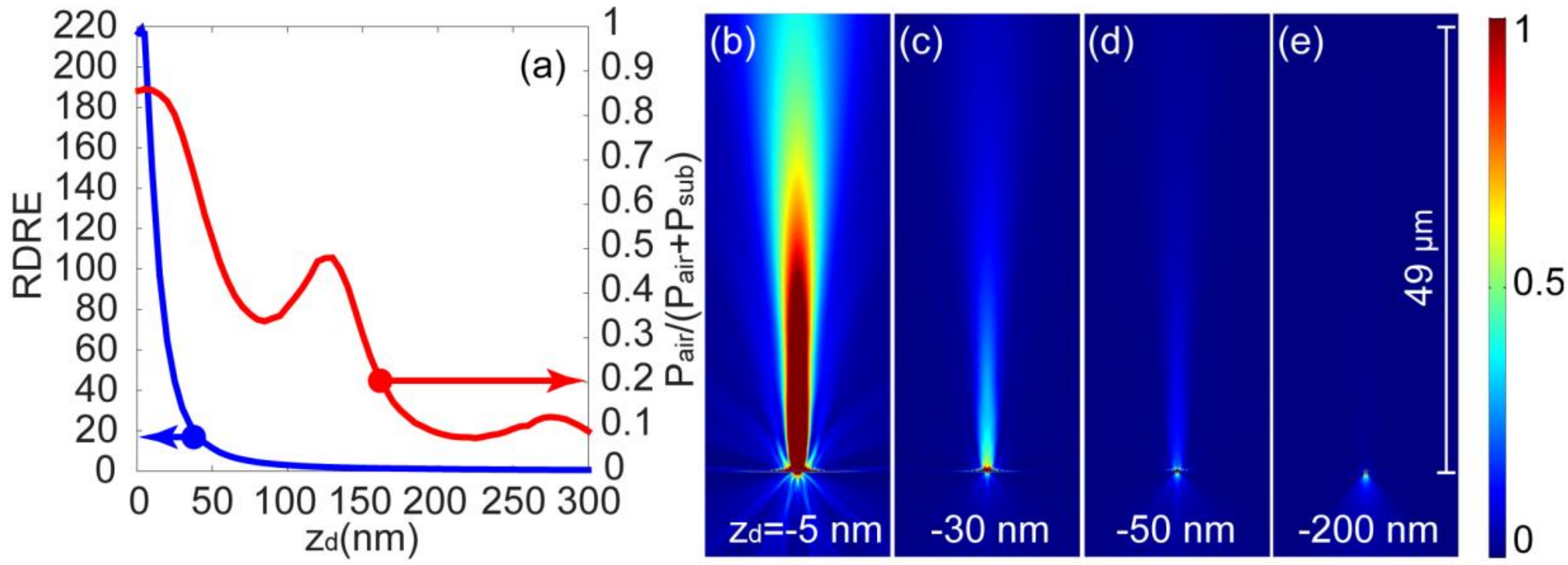

Figure 4: (a) $RDRE_{air}$ and collection efficiency, CE = $P_{air\text{-}C1}/(P_{sub\text{-}C1}+ P_{air\text{-}C1})$ vs. $z_d$. (b)-(e) Radiation patterns, $|E|^2$, for various dipole distances, $z_d$.

For dipole distances $z_d \leq -10$ nm where the near total energy transfer occurs, the donor-acceptor relation between the NV- and the aperture implies that the performance of such a device is sensitive to the donor's dipolar orientation with respect to that of the acceptor. Retaining only the near-field term, radiation pattern of the donor (i.e. NV-) may be written as [21]:

$$E_D(\vartheta) = |\boldsymbol{\mu}_D| \frac{\sin\vartheta}{4\pi\varepsilon_0\varepsilon} \frac{e^{-ikr}}{r^3} \tag{3}$$

where $r = \sqrt{x^2 + z^2}$ with $\vartheta$ being the angle between $r$ and the donor's dipole moment, $\mu_D$, see inset of Figure 5(a). C1's response vs. the direction of $\mu_D$ in the $x$-$z$ plane was calculated in the range of $0° \leq \Theta_{xz} \leq 90°$. Figure 5(a) shows that the power transfer between the donor (NV-) and the acceptor (aperture) does in fact comply with $P_{D\rightarrow A} \propto |\mu_A \cdot E_D|^2$ [21], where $\mu_A$ is the dipole moment of the acceptor. Consequently, for a vertically oriented $\mu_D$, $P_{D\rightarrow A}$ and $RDRE_{air}$ drop to zero. Unlike the simple donor-acceptor interaction observed in florescent molecules, the interaction between the NV- and C1 is influenced by plasmonic effects, therefore surface wave launched by the aperture must also be examined. Numerically obtained phase difference between $E_x$ oscillations on the corners of the aperture at the silver/diamond interface, $\Phi_{\text{L-R}}\big|_{z=0} = \angle E_x(-w_a/2,0) - \angle E_x(w_a/2,0)$, revealed that $\Phi_{\text{L-R}}$ is a function of $\Theta_{xz}$, and may be defined as $\Phi_{\text{L-R}}\big|_{z=0} \equiv f(\Theta_{xz}) \approx 180°\left[1-\cos(\Theta_{xz})\right]$, Figure 5(a). This implies that one can define the parity state of $E_x$ as $\cos(f(\Theta_{xz}))|0\rangle + \sin(f(\Theta_{xz}))|1\rangle$, where $|0\rangle$ and $|1\rangle$ are the states of pure even and pure odd parities with probabilities of $\cos^2(f(\Theta_{xz}))$ and $\sin^2(f(\Theta_{xz}))$ respectively. For pure $|1\rangle$ state of $E_x$, however, one must not assume that NV's emission is quenched. It can be demonstrated that for a vertically oriented $\mu_D$ a small displacement in its horizontal position with respect to the center of the aperture, not only overcomes the suppressed $P_{D\rightarrow A}$, but also causes the transition from the odd to even state, $|1\rangle \rightarrow |0\rangle$, hence satisfying the two mandatory conditions for a high intensity constructive interference along the $z$-axis. Numerical results depicted in Figure 5(b) confirm that for $16 \leq x_d \leq 20$ nm both conditions are satisfied. The trend observed in the $RDRE_{air}$ spectrum is due to the donor's projected field intensity over the aperture, $|E_D|^2_{(x=0,z=0)}$, varying with $x_d$. Analytical values for $|E_D|^2_{(x=0,z=0)}$ vs. $x_d$ were calculated using equation (3), with $\sin\vartheta = |x_d|/r$, see Figure 5(b)-inset. Although, equation (3) does not take into account the aperture's dimension, reflection terms and interactions with surface waves, the trend in $|E_D|^2$ spectrum agrees with that obtained numerically for $RDRE_{air}$. Directional gain of the BE antenna, $|E(\theta)|^2/|E_0(\theta)|^2$, vs. $\Theta_{xz}$ is depicted in Figure 5(c). The field intensity along the optical axis can be 800 times that of a free standing nano-diamond when $\Theta_{xz} = 0°$. Here $|E(\theta)|^2$ and $|E_0(\theta)|^2$ are the electric field intensities as a function of angle, $\theta$, from the $x$-axis calculated for C1 and a nano-diamond particle in vacuum respectively. FWHM of the beam remains approximately 14° for all $\Theta_{xz}$ values. Corresponding directional gains vs. $x_d$ = {10, 16, 21, 33 nm when $\Theta_{xz} = 90°$ are shown in Figure 5(d). The angle $\theta$, may be expressed in terms of $\Theta_{xz}$, $x_d$ and $z_d$ as: $\vartheta = 90° - \left|90° - \Theta_{xz} - \tan^{-1}\left(\frac{|z_d|}{x_d}\right)\right|$, valid in the range of $0° \leq \Theta_{xz} \leq 90°$ and $x_d \geq 0$.

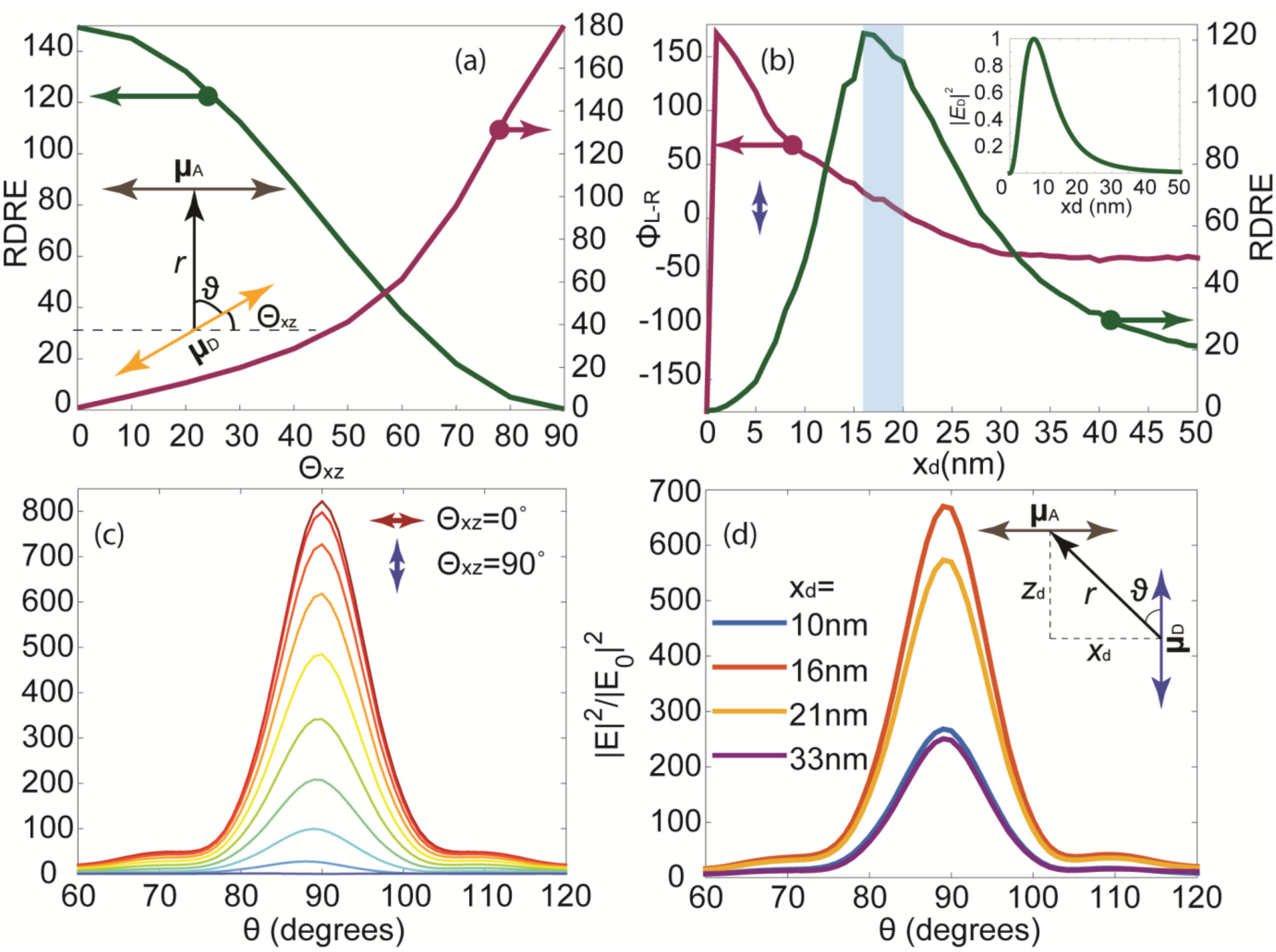

Figure 5. (a) Numerically obtained $\text{RDRE}_{\text{air}}$ and the phase difference between $E_x$ oscillations on the corners of the aperture at the silver/diamond interface, $\Phi_{\text{L-R}}\big|_{z=0} = \angle E_x(-w_a/2,0) - \angle E_x(w_a/2,0)$ vs. the orientation of the $\mu_D$ in the $x$-$z$ plane, $\Theta_{xz}$. (b) Numerically obtained $\text{RDRE}_{\text{air}}$ and $\Phi_{\text{L-R}}$ vs. $x_d$ when $\Theta_{xz} = 90°$. (b)-(inset) Analytical values for donor's field intensity, $|E_D|^2$, projected over the aperture at (0,0) vs. $x_d$ when $\Theta_{xz} = 90°$. (c) Directional gains of the BE antenna, $|E(\theta)|^2/|E_0(\theta)|^2$, for $0° \leq \Theta_{xz} \leq 90°$ when $x_d = 0$. (d) Directional gains for $x_d = \{10, 16, 21, 33\}$ nm when $\Theta_{xz} = 90°$.

Previously, it was shown that the $xz$ (or $yz$) plane of experimentally obtained volumetric radiation patterns of a bullseye structure, corresponds to the radiation pattern of a simulated 2D grating [22](see chapters 8, 9 and Appendix A). To evaluate C1's response vs. the change in $\mu_D$ in the $xy$ plane, C1 was first modelled in 3D as a BE structure with a 250 nm long slit and later with a symmetric cross-shaped aperture having the same arm-lengths. The spherical simulation domain was divided by the planar structure into two hemispheres, with the bottom being the diamond substrate and the upper being the air. The first Fabry-Pérot resonance of the aperture in both cases occurs at $t = 130$ nm. To reduce the required computational resources, number of corrugations were limited to $N = 3$. All other dimensions were kept in accordance to the 2D model. A parametric sweep over the orientation of $\mu_D$ in the $xy$ plane, $\Theta_{xy}$, showed that $\text{RDRE}_{\text{air}} \propto |\mu_A \cdot E_D|^2$ for the slit, as anticipated. $\text{RDRE}_{\text{air}}$ in 3D was calculated using $P_{\text{air}} / P_{\text{sub}} = \int \langle \mathbf{S} \rangle_{\text{air}} \cdot d\mathbf{s} \Big/ \int \frac{\langle \mathbf{S} \rangle_{\text{sub}}}{n_{\text{sub}}} \cdot d\mathbf{s}$ with integrations being carried over the surface $\mathbf{s}$ enclosing the hemisphere. The maximum $\text{RDRE}_{\text{air}}$ obtained from the 3D model was calculated to be ~100 (results not shown here). Far-field intensities, $|E_{\text{far}}|^2$ vs. $\Theta_{xy}$ also followed a similar trend with collimated beams being achieved for $\Theta_{xy} < 90°$, Figure 6(a)-(c). Far-field

intensity emanating from the symmetric cross-shaped aperture, on the other hand, was impervious to the change in $\Theta_{xy}$ but the maximum far-field intensity in this case was almost half of that achieved by the slit, i.e. $|E_{\text{far-cross}}|^2/|E_{\text{far-slit}}|^2 = 0.45$ at their respective maxima. Presence of an additional arm in the cross-shape aperture reduces the nearby metallic surface area over which $|E_D|^2$ is projected on to, hence the reduction in absorption cross section of the acceptor.

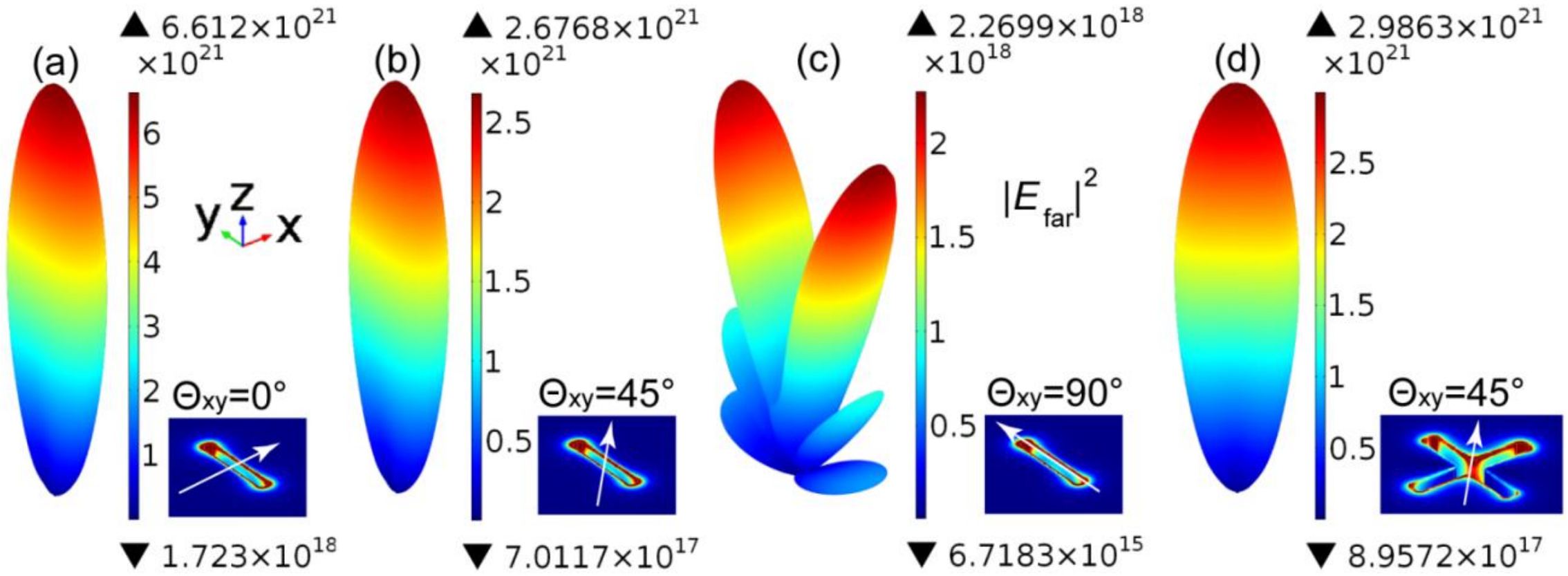

Figure 6. 3D realization of the model with a 30 nm wide and 250 nm long slit and cross-shaped apertures in a 130 nm silver film. Far-field intensities (V/m)² vs. the orientation of $\mu_D$ in the $xy$ plane, $\Theta_{xy}$, for (a)-(c) slit with $\Theta_{xy} = \{0°, 45°, 90°\}$ and (d) cross with $\Theta_{xy} = 45°$.

For a horizontally oriented $\boldsymbol{\mu}_D$, field amplitude of the donor projected over the aperture vs. $x_d$ was calculated analytically using equation (3) with $\sin\vartheta = |z_d|/r$, see Figure 7(a). The trend observed in $|E_D|^2_{(x=0,z=0)}$ spectrum agrees with that of the $\text{RDRE}_{\text{air}}$ obtained numerically, Figure 7(b). Numerical results revealed that the device tolerates a lateral displacement of $\Delta x_d \approx \pm 15$ nm beyond which the $\text{RDRE}_{\text{air}}$ drops below $1/e$ of its maximum. The phase difference $\Delta\Phi_{L\text{-}R}$ in this case rises at $x_d \geq 7$ nm, reaching a maximum of ~105° at $x_d = 23$ nm but never attaining a 180° difference to qualify $E_x$ as a pure odd function. $\text{RDRE}_{\text{air}}$, on the other hand, drops to $1/e$ of it maximum at $x_d \approx 15$ nm where $\Delta\Phi_{L\text{-}R}$ is 70°. Defining the state of even function-ness as $\cos^2(\Delta\Phi_{L\text{-}R}/2)$, a 70° phase difference must drop the $\text{RDRE}_{\text{air}}$ only to 0.67 (not $1/e$) of its maximum. The 55% excess reduction, therefore, is partially due to the $E_x$ not being a pure even function and partially due the reduction in $|E_D|^2_{(x=0,z=0)}$ with increase in $x_d$.

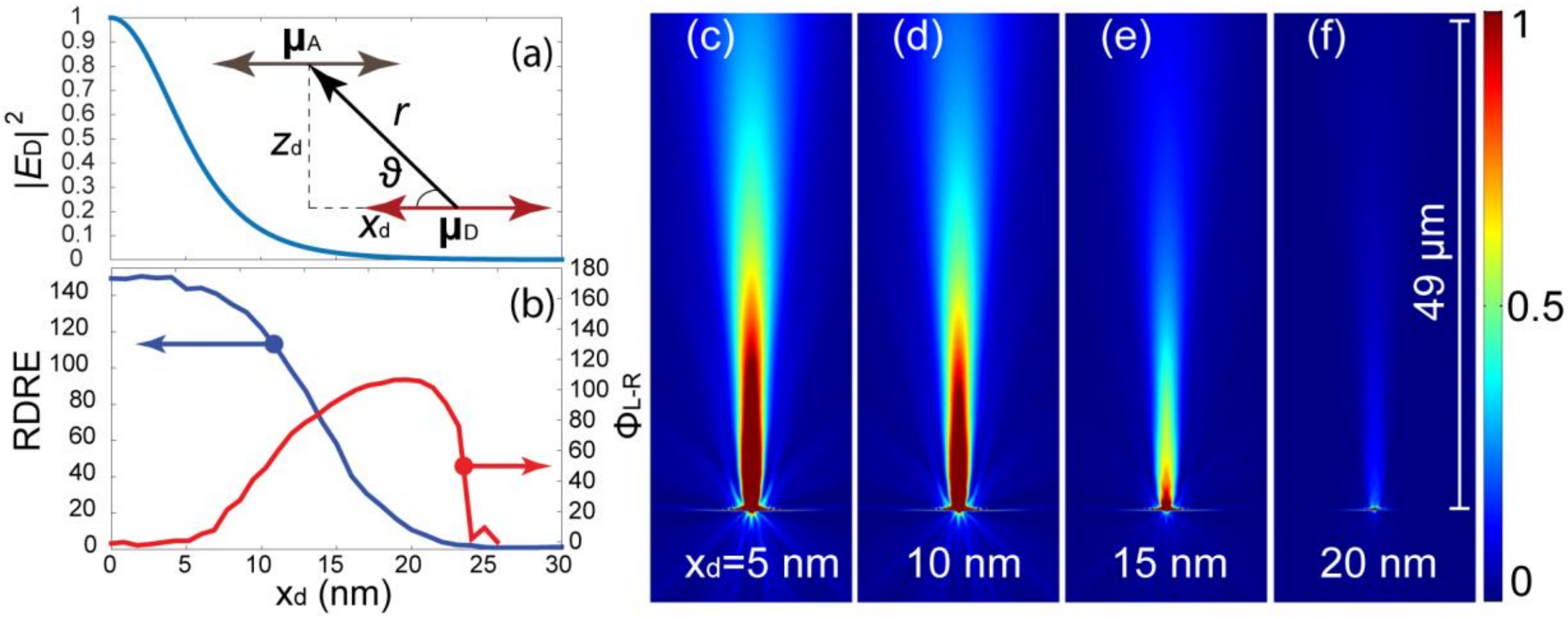

Figure 7. (a) Analytical values for donor's field intensity, $|E_D|^2$, projected over the aperture at (0,0) vs. $x_d$ when $\Theta_{xz} = 0°$. (inset) schematic showing for the donor-acceptor relation. (b) Numerically obtained $\text{RDRE}_{\text{air}}$ and $\Phi_{L\text{-}R}$ vs. $x_d$ when $\Theta_{xz} = 0°$. (c)-(f) Radiation profiles, $|E|^2$, for various values of $x_d$.

## 4. Crystallographic Directions

From all possible dipole orientations considered in this report, it can be inferred that a horizontally oriented $\mu_D$ with a plasmonic slit offers the best outcome with the highest $RDRE_{air}$ and tolerance. Descriptions on triplet notations and Miler indices used in crystallography, may be found in undergraduate solid state texts[45,46]. With NV- symmetry axis being one of the <111> crystallography directions[26,27], one may narrow the search for a horizontally oriented $\mu_D$. Assuming that the planar design lies on an arbitrary plane, z, defined by its normal vector $\hat{z}$ , possible scenarios are (a) $z$ plane parallel to diamond's $(1\bar{1}0)$ plane which implies that $\mu_D$ is parallel to the $[\bar{1}\bar{1}2]$ direction, hence the slit must be fabricated such that $\mu_A \parallel \mu_D$. To maximize the emission from the NV- in this case, the excitation field, $\mathbf{E}_{ex}$, must also be polarized in the $[\bar{1}\bar{1}2]$ direction. (b) $z$ plane $\parallel (\bar{1}\bar{1}2)$ with $\mathbf{E}_{ex} \parallel \mu_A \parallel \mu_D \parallel [1\bar{1}0]$. (c) $z$ plane‖111) (or $\hat{z}$ being aligned with the symmetry axis of the NV-) with no preferential direction for neither $\mathbf{E}_{ex}$ nor $\mu_D$ since $\mu_D = \alpha^2\mu_{D1} + \beta^2\mu_{D2}$, where $\alpha^2 + \beta^2 = 1$, see Figure 8. Depending on the application, one can identify other diamond planes over which the planar structure to be set. For examples: if the emphasis is on the extraction of single photons, then the diamond plane must be chosen such that ($\mu_A \parallel \mu_{D1}$ AND $\mu_A \perp \mu_{D2}$) OR ($\mu_A \parallel \mu_{D2}$ AND $\mu_A \perp \mu_{D1}$). On the other hand, if one is to investigate the two-entangled photon arising from the mutual excitations of $\mu_{D1}$ and $\mu_{D2}$, then it is best to implement the planar structure using a symmetric cross aperture on the (111) plane with the arms of the cross being aligned with $[\bar{1}\bar{1}2]$ and $[1\bar{1}0]$.

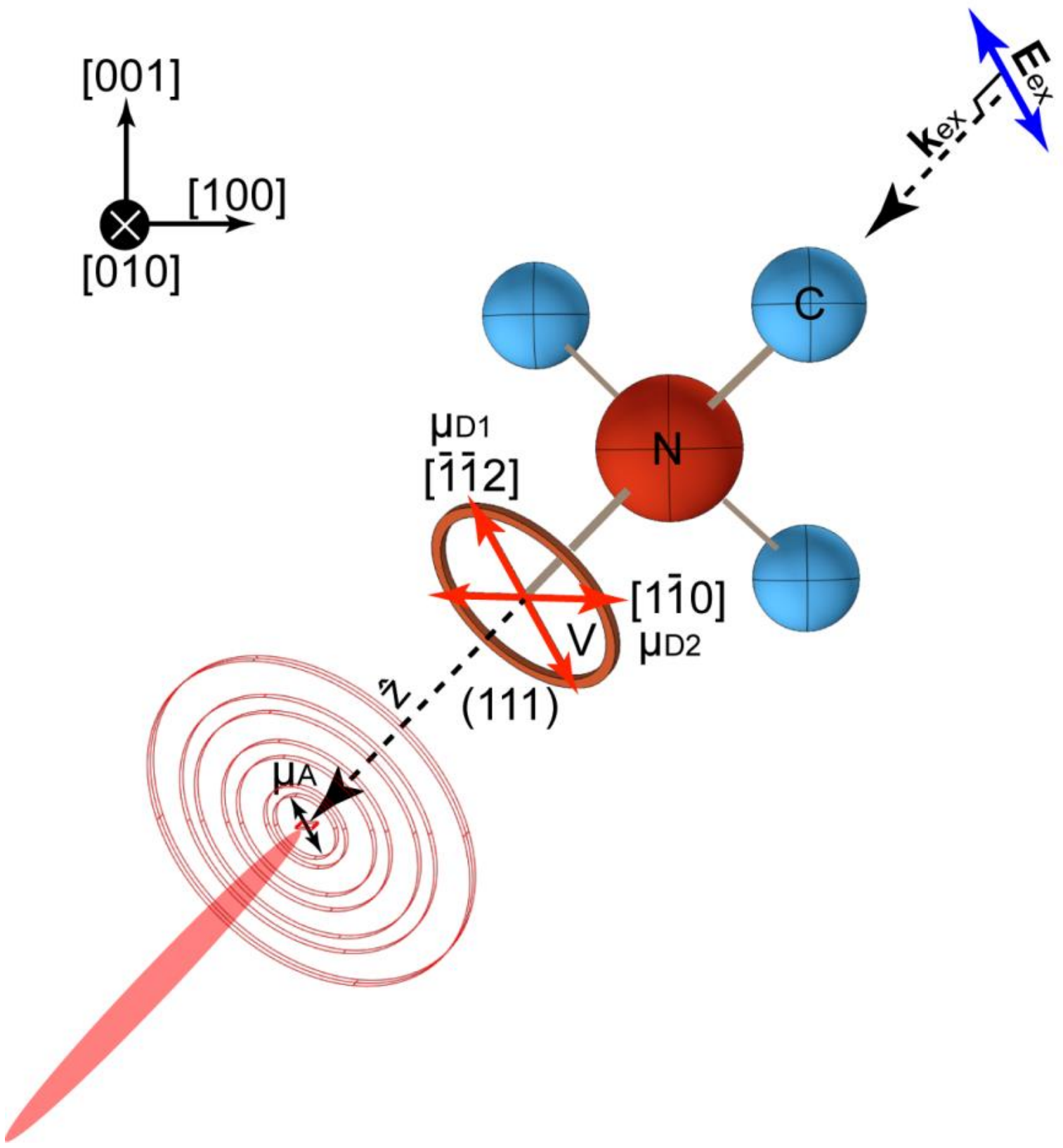

Figure 8: Best case scenario for implementing $\mu_A$ with respect to $\mu_D$ with $z$ plane||(111) (or $z$-axis being aligned with the symmetry axis of the NV-). There is no preferential direction for either $\mathbf{E}_{ex}$ or $\mu_D$ since $\mu_D = \alpha^2\mu_{D1} + \beta^2\mu_{D2}$, where $\alpha^2 + \beta^2 = 1$.

## 5. Conclusion

Calculations demonstrated the donor-acceptor relation between a quantum emitter and a resonant aperture in Surface Plasmon-Coupled Enhanced Transmission (SPCET). The emission of a dipole captured by and transmitted through a metallic resonant aperture is reciprocally related to the quenching of dipole's emission near flat metallic surface. Meaning that, for distances where the dipole emission is quenched near a flat metallic surface, transmission through a metallic resonant aperture is maximized. Unidirectional scattering of SPCET through the aperture into the air is particularly advantageous in moving away from designs based on the reflection geometry that utilize epi-illumination/collection from optically dense substrates or superstrates. Considering that SPP propagation over the plasmonic bullseye structure involves Fabry-Pérot-like resonances between the grooves and the aperture, a new approach for designing the corrugations was proposed with periodicity that differs significantly from the SPP wavelength. Combining the SPCET effect with the newly proposed BE structure, it was shown that it is possible to collimate the radiation of a quantum emitter (positioned inside the substrate) with a high yield and collection efficiency in the air. The platform was studied with diamond substrate as the material of choice with NV- near the surface. The dipole moment of a NV- in a single crystal diamond retains its direction at all time. By characterizing potential NV- centers and their dipolar orientations in a single crystal diamond substrate, it is then possible to fabricate the structure over an already identified NV- to maximize the overall performance of the device. Depending on the application, the planar plasmonic structure may be fabricated over a specific diamond plane. The challenge, however, is to grow the diamond film so that the target plane is exposed to the proposed plasmonic structure.

**Conflicts of Interest:** The author declares no conflict of interest. In the case of the computational resources availed to the author being interpreted as fund of some sort, the author wishes to state that the founding sponsors had no role in the inception of the idea, the design of the study; in the collection, analyses, or interpretation of data; in the writing of the manuscript, and in the decision to publish the results.

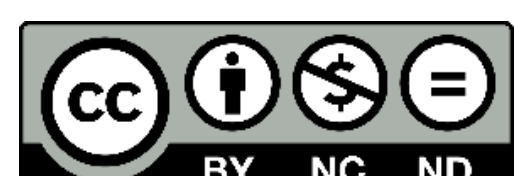